\input lanlmac
\input epsf.tex
\overfullrule=0pt

\newcount\figno
\figno=0
\def\fig#1#2#3{
\par\begingroup\parindent=0pt\leftskip=1cm\rightskip=1cm\parindent=0pt
\baselineskip=11pt
\global\advance\figno by 1
\midinsert
\epsfxsize=#3
\centerline{\epsfbox{#2}}
\vskip 12pt
{\bf Fig.\ \the\figno:} #1\par
\endinsert\endgroup\par
}
\def\figlabel#1{\xdef#1{\the\figno}%
\writedef{#1\leftbracket \the\figno}%
}
\def\omit#1{}

\def\pre#1{{\tt
#1}}%use this to give preprint # in refs

\def\Rc{{\check R}}
\def\pf{{\rm Pf}}

\def\qed{\nobreak\hfill\vbox{\hrule height.4pt%
\hbox{\vrule width.4pt height3pt \kern3pt\vrule width.4pt}\hrule height.4pt}\medskip\goodbreak}
% References
\lref\RS{A.V. Razumov and Yu.G. Stroganov, 
{\sl Combinatorial nature
of ground state vector of $O(1)$ loop model},
{\it Theor. Math. Phys.} 
{\bf 138} (2004) 333--337; {\it Teor. Mat. Fiz.} 138 (2004) 395--400, \pre{math.CO/0104216}.}
\lref\BdGN{M.T. Batchelor, J. de Gier and B. Nienhuis,
{\sl The quantum symmetric XXZ chain at $\Delta=-1/2$, alternating sign matrices and 
plane partitions},
{\it J. Phys.} A34 (2001) L265--L270,
\pre{cond-mat/0101385}.}
\lref\IZER{A. Izergin, {\sl Partition function of the six-vertex
model in a finite volume}, {\it Sov. Phys. Dokl.} {\bf 32} (1987) 878--879.}
\lref\KOR{V. Korepin, {\sl Calculation of norms of Bethe wave functions},
{\it Comm. Math. Phys.} {\bf 86} (1982) 391--418.}
\lref\FR{I.B.~Frenkel and N.~Reshetikhin, {\sl Quantum affine Algebras and Holonomic 
Difference Equations},
{\it Commun. Math. Phys.} 146 (1992), 1--60.}
\lref\DFZJ{P.~Di Francesco and P.~Zinn-Justin, {\sl Around the Razumov--Stroganov conjecture:
proof of a multi-parameter sum rule}, {\it Elec. Jour. Comb.} 12 (1) (2005), R6,
\pre{math-ph/0410061}.}
\lref\DFOP{P. Di Francesco, {\sl Inhomogenous loop models with open boundaries},
J. Phys. A: Math. Gen. {\bf 38} (2005) 6091--6120, \pre{math-ph/0504032}}
\lref\Pas{V.~Pasquier, {\sl Quantum incompressibility and Razumov Stroganov type conjectures},
\pre{cond-mat/0506075}.}
\lref\ZJDF{P.~Di Francesco and P.~Zinn-Justin, {\sl Quantum Knizhnik--Zamolodchikov equation,
generalized Razumov--Stroganov sum rules and extended Joseph polynomials}, 
{\it J. Phys. A } {\bf 38}
(2006)  L815--L822, \pre{math-ph/0508059}.}
\lref\DF{P.~Di Francesco, {\sl A refined Razumov--Stroganov conjecture},
J. Stat. Mech. P08009 (2004), \pre{cond-mat/0407477}.}
\lref\STEM{J. Stembridge, {\sl Nonintersecting paths, Pfaffians and plane partitions},
Advances in Math. {\bf 83} (1990) 96--131.}
\lref\Bre{D. Bressoud, {\sl Proofs and Confirmations: The Story of
the Alternating Sign Matrix Conjecture}, Cambridge Univ. Pr., 1999.}
\lref\DFZJZ{P. Di Francesco, P. Zinn-Justin and J.-B. Zuber,
{\sl Sum rules for the ground states of the $O(1)$ loop model
on a cylinder}, 
J. Stat. Mech.: Theor. Exp. (2006) P08011, \pre{math-ph/0603009}.}
\lref\ORBI{P. Di Francesco and P. Zinn-Justin,{\sl From Orbital Varieties to Alternating Sign Matrices},
submitted to FPSAC '06, \pre{math-ph/0512047}.}
\lref\Ku{G. Kuperberg, {\sl Symmetry classes of alternating-sign
matrices under one roof}, 
{\it Ann. of Math.} 156 (3) (2002), 835--866,
\pre{math.CO/0008184}.}
\lref\DFTS{P.~Di Francesco, {\sl Totally Symmetric Self Complementary Plane Partitions
and the quantum Knizhnik-Zamolodchikov equation: a conjecture},
J. Stat. Mech.: Theor. Exp. (2006) P09008,  \pre{cond-mat/0607499}.}
\lref\RStr{A.V. Razumov and Yu.G. Stroganov, 
{\sl $O(1)$ loop model with different boundary conditions and symmetry classes 
of alternating-sign matrices},
{\it Theor. Math. Phys.} 
{\bf 142} (2005) 237--243; {\it Teor. Mat. Fiz.} 142 (2005) 284--292,
\pre{cond-mat/0108103}.}
\lref\PRdG{P. Pearce, V. Rittenberg and J. de Gier,{\sl Critical Q=1 Potts model
and Temperley--Lieb stochastic processes}, \pre{cond-mat/0108051}; P. Pearce, V. Rittenberg, 
J. de Gier and B.~Nienhuis, 
{\sl Temperley--Lieb Stochastic Processes},
J. Phys. {\bf A35} (2002) L661--L668,
\pre{math-ph/0209017}.}
\lref\MRR{W. Mills, D. Robbins and H. Rumsey, {\sl Enumeration of a symmetry class of 
plane partitions}, Discrete Math. {\bf 67} (1987) 43--55.}
\lref\LGV{B. Lindstr\"om, {\it On the vector representations of
induced matroids}, Bull. London Math. Soc. {\bf 5} (1973)
85--90\semi
I. M. Gessel and X. Viennot, {\it Binomial determinants, paths and
hook formulae}, Adv. Math. { \bf 58} (1985) 300--321. }
\lref\DFZJb{P.~Di Francesco and P.~Zinn-Justin, {\sl Inhomogeneous model of crossing loops
and multidegrees of some algebraic varieties}, 
to appear in {\it Commun. Math. Phys.} (2005),\break
\pre{math-ph/0412031}.}
\lref\KZJ{A. Knutson and P. Zinn-Justin, {\sl A scheme related to the Brauer loop model}, \pre{math.AG/0503224}.}
\lref\KM{A.~Knutson and E.~Miller, {\sl Gr\"obner geometry of Schubert polynomials},
{\it Annals of Mathematics} (2003), \pre{math.AG/0110058}.}
\lref\CK{M. Ciucu and K. Krattenthaler, {\sl Plane partitions: $5{1\over 2}$ symmetry classes},
in "Combinatorial Methods in Representation Theory," 
(M. Kashiwara, K. Koike, S. Okada, I. Terada, and
H. Yamada, Eds.), Advanced Studies in Pure Mathematics 
{\bf 28} (2000) 81--103, \pre{math.CO/9808018}.}
\lref\KK{C. Krattenthaler,
{\sl Advanced determinant calculus}, S\'eminaire Lotharingien Combin. ("The Andrews Festschrift")
{\bf 42} (1999) B42q.}
\lref\CEKZ{M. Ciucu, T. Eisenk\"olbl, C. Krattenthaler
and D. Zare, {\sl Enumeration of lozenge tilings of hexagons with a central triangular hole},
J. Combin. Theory Ser. A {\bf 95}(2001) 251-334, \pre{math.CO/9912053};
C. Krattenthaler, {\sl Descending plane partitions and rhombus tilings of a
hexagon with triangular hole}, Europ. J. Combin. {\bf 27} (2006) 1138-1146, \pre{math.CO/0310188}.}
\lref\KAPA{M. Kasatani and V. Pasquier, {\sl On polynomials interpolating between the stationary 
state of a O(n) model and a Q.H.E. ground state}, \pre{cond-mat/0608160}.}

\Title{SPhT-T06/144}
{\vbox{
\centerline{Open boundary Quantum}
\medskip
\centerline{Knizhnik-Zamolodchikov equation}
\medskip
\centerline{and the weighted enumeration} 
\medskip
\centerline{of Plane Partitions with symmetries}
}}
\bigskip\bigskip
\centerline{P.~Di~Francesco} 
\medskip
\centerline{\it  Service de Physique Th\'eorique de Saclay,}
\centerline{\it CEA/DSM/SPhT, URA 2306 du CNRS,}
\centerline{\it F-91191 Gif sur Yvette Cedex, France}
\bigskip
\vskip0.5cm
%abstract
\noindent We propose new conjectures relating sum rules for the polynomial solution of the 
qKZ equation with open (reflecting) boundaries as a function of the quantum parameter $q$
and the $\tau$-enumeration of Plane Partitions with specific symmetries, 
with $\tau=-(q+q^{-1})$. We also find a conjectural relation \`a la Razumov-Stroganov
between the $\tau\to 0$ limit of the qKZ solution and refined numbers of
Totally Symmetric Self Complementary Plane Partitions.

\bigskip

AMS Subject Classification (2000): Primary 05A19; Secondary 82B20
%\draft
\Date{01/2007}
%
%%%%%%%%%%%%%%%%%%%%%%%%%%%%%%%%%%%%%%%%%%%%%%%%%%%%%%%%%%%%%%%%%%%%%
%

\newsec{Introduction}

Integrable lattice models seem to be a constant source of combinatorial
wonders. Any statistical lattice model is combinatorial by essence, as
it is based on the (weighted) enumeration of configurations. Integrability
appears then as the ``cherry on the cake" that gives access to exact solutions
and, from a purely combinatorial point of view, to exact and/or asymptotic enumeration,
involving the analytical computation of critical configuration exponents.

This note is devoted to an extension of the so-called Razumov-Stroganov (RS)
conjecture \RS, identifying the properly normalized entries of the groundstate (Perron-Frobenius)
eigenvector $\Psi$ of the O(1) 
dense loop model on a cylinder of perimeter $2n$ in the basis of link patterns with
the numbers of configurations of the Fully Packed Loop (FPL) model on an $n\times n$ square grid,
corresponding to the {\it same} link patterns. A weaker ``sum rule" version \BdGN\ of this conjecture
simply states that the sum of the components of $\Psi$ equals the total number of FPLs,
itself equal to that of Alternating Sign Matrices (ASMs) of the same size. The latter was 
first proved in \DFZJ, by making extensive use of the integrability of a more general
inhomogeneous version of the O(1) model. There, it is shown that $\Psi$ may be entirely
determined by translating any permutation of the inhomogeneity parameters (spectral parameters)
in terms of the local action of the Temperley-Lieb algebra generators, resulting into
divided difference equations obeyed by the components of $\Psi$, that are
homogeneous polynomials of the spectral parameters, tending to the above integers in the
``homogeneous" limit where all spectral parameters tend to $1$. In \DFZJ, the
sum of components of $\Psi$ is actually computed and identified with a particular
case of the Izergin-Korepin determinant \IZER\ \KOR, reducing to a simple Schur function in that case.

This remarkable link between the O(1) integrable model and ASMs adds up yet another piece to the
long lasting puzzle of the Alternating Sign Matrices (see Bressoud's book \Bre\ for a thrilling tale).
ASMs indeed seem to be mysteriously related to other combinatorial objects such as Descending Plane
Partitions, and even more interestingly to Plane Partitions with specific symmetries. The latter
may all be viewed as rhombus tilings of various domains of the triangular lattice, by means
of elementary rhombi made of two adjacent triangles.
Particularly interesting are the Totally Symmetric Self Complementary Plane Partitions
(TSSCPPs), which may be viewed as rhombus tilings of a regular hexagon of size 
$(2n)\times (2n)\times (2n)$,
and which moreover enjoy all possible symmetries of the hexagon. 
A Plane Partition is indeed a 
pile of elementary cubes inside a cube of size $(2n)\times (2n)\times (2n)$. When 
viewed in perspective from the $(1,1,1)$ direction, the visible individual cube tops and sides
form rhombi which tile the large cube's projection, a regular hexagon of size $2n$. 
The desired symmetries of the pile of cubes, namely that under rotations of
$2\pi/3$ around the axis $(1,1,1)$ and the self-complementation meaning that the
complement of the pile within the large cube is itlelf a pile with the same structure,
translate into a maximal symmetry of the hexagon's rhombus tilings.
Although no canonical bijection is known to this day between TSSCPPs
and ASMs, their numbers are identical.

Razumov and Stroganov also considered the O(1) loop model on a strip of width $L$
rather than on a cylinder \RStr, thus trading {\it periodic} for {\it open} boundary conditions, 
and identified again the properly normalized components of the corresponding groundstate 
vector with the 
numbers of Vertically Symmetric FPLs (VSFPLs), that is FPLs that are reflection-symmetric
with respect to a vertical axis, themselves identified with Vertically Symmetric ASMs (VSASMs) for
even size $L$. Similarly, for odd size $L$, the sum of components of the properly normalized
groundstate vector was conjectured in \PRdG\ to be given by the number
of Cyclically Symmetric Transpose Complement Plane Partitions (CSTCPPs). 
The latter Plane Partitions 
enjoy cyclic rotational symmetry under rotations of $2\pi/3$, and are moreover 
equal to the complement of their reflection. 
The sum rules for the open boundaries were computed in \DFOP, along the same
lines as \DFZJ, resulting in simple determinant and
Pfaffian expressions in terms of spectral parameters. From the result of \DFOP, it is 
a simple exercise to compute the homogeneous limit of the sum rule, which reduces for $L=2n$ to
$A_V(2n+1)$, the total number of VSASMs of size $2n+1$, and for $L=2n-1$ to 
$N_8(2n)$, the total number of CSTCPPs of a regular hexagon of size $(2n)\times (2n)\times (2n)$. 
The duplicity of this result makes one think that the language of
Plane Partitions might also be useful to approach the RS conjectures.

An alternative subsequent proof of the periodic boundary RS sum rule
may be found in \Pas, where the integrability of
the model is put into perspective within the framework of the Affine Temperley-Lieb 
algebra and its representation theory. A further reformulation in terms
of the $U_q(sl_2)$ quantum Knizhnik-Zamolodchikov (qKZ) equation has led to a host 
of generalizations, both to higher rank algebras \ZJDF\ and 
to different boundary conditions \ORBI\ 
(indexed by root systems of classical Lie algebras).
All these extensions involve an extra (quantum group) parameter $q$, 
equal to $-e^{i\pi/3}$ in the RS case, via the quantity
\eqn\deftau{ \tau=-q -{1\over q} }
This is nothing but the weight per loop one would assign within the Temperley-Lieb algebra framework,
when dealing with the more general $O(n=\tau)$ model, however when $\tau\ne 1$ (i.e. except at the RS
point), no nice cylinder partition function interpretation holds: indeed, for generic $q$, 
the boundary conditions are not {\it periodic}, but only {\it cyclic} up to a 
multiplicative shift of $q^6$ on the spectral parameters, $z_{i+L}\to q^6 z_i$.

Nevertheless, after taking the homogeneous limit,
the solution to the cyclic qKZ equation now produces
a vector $\Psi(\tau)$ whose properly normalized components are polynomials of $\tau$, with 
apparently {\it non-negative integer} coefficients.
In Ref.\DFTS, we have identified the sum rule for the components of $\Psi(\tau)$
with the weighted $\tau$-enumeration of TSSCPPs, carrying a weight $\tau$ per 
vertical step in their Non-Intersecting Lattice Path (NILP)
formulation. The latter is a reexpression of the TSSCPPs in a fundamental domain of the hexagon
($1/12$-th of it) in terms of lattice paths drawn on the rhombi, that do not intersect. The
counting of such paths is now a standard exercise. Ref.\DFTS\ therefore provides a conjectural 
combinatorial interpretation for the non-negative integer coefficients of $\Psi(\tau)$ 
(these non-negative integers were also spotted in \KAPA, but without combinatorial interpretation).

The aim of this paper was to extend the conjecture of \DFTS\ to the case of open boundary conditions.
On the way, we have found a remarkable coincidence between the leading coefficient of the entries
$\Psi_\pi(\tau)$ of the open qKZ solution $\Psi(\tau)$ at small $\tau$ and the refined TSSCPP numbers,
that count TSSCPPs grouped according to the positions of their endpoints in the NILP formulation.
Our next task, to restore symmetry between the cases of even and odd strip width,
was to find some Plane Partition interpretation of the VSASM numbers, and we found
out that the latter also count rhombus tilings of a hexagon with the same symmetries as for the CSTCPPs,
but now with a central triangular hole of size $2\times 2\times 2$. Note that this 
hexagon is no longer regular, but with shape $(2n)\times (2n+2)\times (2n)$. Note also that this
allows for a unified NILP interpretation of both VSASM and CSTCPP numbers.

This led us to the main conjectures of this paper: 
\item{(i)} {\it the leading terms in the components of the qKZ solution $\Psi(\tau)$ when $\tau\to 0$
are the refined TSSCPP numbers arranged according to the endpoints of their associated NILP, with a
simple bijection relating these to link patterns.}
\item{(ii)}{\it the sum rule for the properly normalized solution
$\Psi(\tau)$ of the $U_q(sl_2)$ qKZ equation with open boundaries produces a polynomial of $\tau$
with non-negative integer coefficients, identical to the generating polynomials for VSASMs 
(of size $2n+1$ for $L=2n$)
and CSTCPPs (of size $2n$ for $L=2n-1$) with a weight $\tau$ per vertical step 
in their respective NILP formulations,
except for the steps in one particular central row.}
\item{(iii)}{\it the ``maximal" components of $\Psi(\tau)$ corresponding to the link pattern that connects
all points $2i-1$ to $2i$, leaving the point $L$ unmatched in the odd case, are nothing 
but the generating polynomials for CSSTPPs (of size $2n-2$ for $L=2n$) and VSASMs 
(of size $2n-1$ for $L=2n-1$) with a weight $\tau$ per vertical step in their respective 
NILP formulations, and without any further restriction.}
\par
\noindent While the second conjecture is only a sum rule, the first one, like the full RS
conjecture, involves separately each component of $\Psi(\tau)$ and gives a combinatorial 
interpretation for the leading term when $\tau\to 0$ in terms of TSSCPPs rather than ASMs or FPLs
involved in the RS conjecture. This change of point of view, trading ASMs or FPLs for TSSCPPs should be
very fruitful, and suggests that the O(1) loop model or the qKZ solutions may be the right place 
where to look for some ASM/FPL - TSSCPP correspondence.

The paper is organized as follows. In Section 2, we recall some known facts on
the qKZ equation with open boundaries and its minimal polynomial solution, which we list for
sizes up to $L=8$ in Appendix A, in their homogeneous form, that keeps only the $\tau$ dependence.
Section 3 reviews Plane Partitions with various symmetries, namely TSSCPPs and CSTCPPs, and introduces
a rhombus tiling problem whose count matches the number of VSASMs, 
and provides a natural generalization of CSTCPPs in the case of even size $L$. In Section 4, we
introduce the $\tau$-enumeration of CSTCPPs and their even counterparts, which eventually match 
the sum rules for the homogeneous solutions of the open qKZ equation in odd and even size,
listed in Appendix A. Section 5 gathers the various conjectures of the paper, as well as some 
concluding remarks.

\newsec{qKZ equation with open boundaries}

\subsec{The equation}

\fig{A sample link pattern $\pi$ in size $L=10$ (a) and the associated Dyck 
path $p(\pi)$ of length $10$ (b). We have also indicated the box decomposition of the Dyck path,
having $\beta(\pi)=4$ boxes here.}{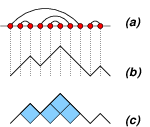}{6.cm}
\figlabel\lipadyck

We refer to \DFOP\ for a detailed presentation. Let us introduce link patterns of size $L$
as configurations of $L$ regularly spaced points on a line, labelled $1$ to $L$ from left to right,
and connected by pairs through 
non-intersecting semi-circles with centers on the line, all contained in the 
upper-half plane delimited by the line. For even $L$, all points are matched, while for
odd $L$, one of them remains unmatched, and should be thought of as connected to the infinity 
on the strip via an infinite half-line not crossing any semi-circle (we'll drop this half-line
for simplicity in pictorial representations). 
There are $c_n=(2n)!/(n!(n+1)!)$ such link patterns
for $L=2n$ and $L=2n-1$, the $L=2n-1$ link patterns being in bijection with that at $L=2n$ 
upon sending the point labelled $L$ to infinity.
A standard bijection replaces link patterns by Dyck paths (see Fig.\lipadyck), 
namely lattice paths of $L$ steps,
starting from the origin of the integer plane, making steps $(1,1)$ or $(1,-1)$ only, 
visiting only points $(x,y)$ with $y\geq 0$ and ending at point $(L,0)$ if $L$ is even, 
and $(L,1)$
if $L$ is odd. Visiting the link pattern $\pi$ from left to right, we define the Dyck path $p(\pi)$
as follows: it takes an $i$-th step $(1,1)$ (resp. $(1,-1)$) if a semi-circle originates 
(resp. terminates) at point $i$ on $\pi$, and an $i$-th step $(1,1)$ if the point $i$ is unmatched
in $\pi$. A useful notion is that of box decomposition of the path, namely expressing it as the
hull of the pile of tilted squares of size $\sqrt{2}$ on top of a zig-zag line between 
the lines $y=0$ and $y=1$ (see Fig.\lipadyck\ for an example). We denote by $\beta(\pi)$
the number of boxes in the decomposition of $p(\pi)$.

There is a natural action of the Temperley-Lieb algebra $TL(\tau)$ on link patterns. The generators
$e_i$, $i=1,2,\ldots,L-1$ act by inserting a small semi-circle connecting points $i$ and $i+1$,
while gluing the former arcs issued from $i$ and $i+1$ into a single arch connecting 
their respective other ends. If $i$ or $i+1$ is unmatched, $e_i$
simply switches $i$ and $i+1$ (i.e. interchanges the positions of the unmatched point and of the
matched one). In the case where $i$ and $i+1$ are already
connected, the link pattern is left unchanged, but receives a multiplicative factor $\tau$.
This leads to the celebrated Temperley-Lieb algebra relations; $e_i^2=\tau e_i$ and
$e_ie_{i\pm 1} e_i=e_i$.

The dense loop model with weight $\tau=-(q+q^{-1})$ per loop is defined via the $R$-matrix
\eqn\rmat{\Rc_{i,i+1}(z,w)= {q^{-1} z -qw \over q^{-1} w -q z} I +{z-w\over q^{-1} w -q z} e_i}
which we may view as an operator acting on link patterns, $I$ acting as the identity.
Let us denote by $\tau_i$ the operator that interchanges $z_i\leftrightarrow z_{i+1}$
in any function of the parameters $z_1,z_2,\ldots, z_L$.
The level one $U_q(sl_2)$ qKZ equation with open boundaries reduces to the system:
\eqn\sysopqkz{
\eqalign{
\tau_i\Psi(z_1,\ldots,z_L)&=\Rc_{i,i+1}(z_{i+1},z_i)\Psi(z_1,\ldots,z_L), 
\quad  1\leq i\leq L-1\cr
\Psi({r\over z_1},z_2,...,z_{L-1},z_L)&= c_1(z_1) \Psi(z_1,...,z_L)\cr
\Psi(z_1,z_2,...,z_{L-1},{rs\over z_L})&= c_L(z_L) \Psi(z_1,...,z_L)\cr
}}
where $s=q^6$ and $c_1$ and $c_L$ two functions to be determined, and $\Psi$ 
a vector in the link pattern basis. 
In the following we will restrict ourselves to the values $r=1$, $rs=q^6$, of the 
boundary terms\foot{It seems that only the cases $r=1$ and $r=1/q^6$ produce nice polynomals
of $\tau$ with integer coefficients for the components of $\Psi$, the two being interchanged
under the reflection of link patterns with respect to a vertical line.}.

Using the expression for $\Rc_{i,i+1}$ \rmat, we may rewrite the first equation of \sysopqkz\
above in components (indexed by link patterns $\pi$) as:
\eqn\compopsi{
\Delta_i\Psi_\pi=\sum_{\pi'\neq \pi\atop e_i\pi'=\pi} \Psi_{\pi'}, \qquad 1\leq i\leq L-1}
where the operator $\Delta_i$ reads
\eqn\deldef{ \Delta_i= {q^{-1}z_{i+1}-qz_i\over z_{i}-z_{i+1}} (\tau_i -1) }

In \DFOP, it was shown that it is sufficient to solve these equations in the case 
of even size $L=2n$, as the solution for size $L-1$ may then be obtained by taking 
the limit $z_L\to 0$, namely
\eqn\solminone{ \Psi_{f(\pi)}(z_1,\ldots,z_{L-1})\propto  \Psi_{\pi}(z_1,\ldots,z_{L-1},0)}
while the link patterns are mapped bijectively $\pi\to f(\pi)$ by removing the point $L$ and leaving
unmatched the point formerly connected to it. 
Unless otherwise stated, we restrict ourselves to $L=2n$ in the following. 

\subsec{Minimal polynomial solution $\Psi$}

In Ref.\DFOP\ it was argued that the minimal polynomial solution to the qKZ equation
has the following basic component corresponding to the fully nested link pattern
$\pi_0$ that connects points $i$ to $2n+1-i$, $i=1,2,\ldots,n$:
\eqn\funda{ \Psi_{\pi_0}=\prod_{1\leq i<j\leq n} (qz_i-q^{-1}z_j)(q-q^{-1}z_iz_j)
\prod_{n+1\leq i<j\leq 2n}(qz_i-q^{-1}z_j)(q^{-2}z_iz_j-q^{2})}
which clearly satisfies the boundary conditions of Eq.~\sysopqkz\ with $c_1(x)=1/x^{2n-2}$
and $c_L(x)=(q^3/x)^{2n-2}$. 
Then, as explained in Ref.\DFOP, using Eq.~\compopsi, all other components of $\Psi$ are expressed 
in a triangular way as linear combinations of products of operators $\Delta$ acting on $\Psi_{\pi_0}$.

The first few solutions for $L=1,2,\ldots,8$ are given in Appendix A below for completeness, in the
homogeneous limit where all $z_i\to 1$ (for even size $L$), except for $z_L\to 0$ (for odd size $L-1$), 
and upon dividing out by a global factor 
$(q-q^{-1})^{2n(n-1)}$ for even size $L$ and $(-q)^{3n-3}(q-q^{-1})^{2(n-1)^2}$ for odd size $L-1$, 
and using the variable $\tau$ of Eq.~\deftau. 
We define the sum rule $\Pi_L(\tau)$ to be simply the sum of components of $\Psi(\tau)$ normalized
in this way.

\subsec{Miscellaneous conjectures}

Inspecting the examples of Appendix A, we have come up with a few conjectures that we list below.

\noindent{\bf Degree and valuation:}
We note the following pattern
for the degree and valuation (highest and lowest powers of $\tau$) 
of $\Psi_\pi(\tau)$ as a polynomial of $\tau$.
Expressing the link patterns as Dyck paths, recall that $\beta(\pi)$ denotes the
number of boxes in the decomposition of $p(\pi)$. In the odd case
$L=2n-1$, let us also record the position $u(\pi)$ of the unmatched point in $\pi$, 
$u(\pi)=1,3,5,\ldots 2n-1$. 
Then we have:
\eqn\valdeg{\eqalign{ {\rm deg}(\Psi_\pi)&= n(n-1)-\beta(\pi) \qquad {\rm for}\ L=2n \cr
{\rm deg}(\Psi_\pi)&= (n-1)^2-\beta(\pi) \qquad  \, {\rm for} \ L=2n-1\cr
{\rm val}(\Psi_\pi)&= \beta(\pi)\qquad \qquad\qquad \ \ \ {\rm for}\ L=2n\cr 
{\rm val}(\Psi_\pi)&= \beta(\pi)+u(\pi)-n \qquad {\rm for}\ L=2n-1\cr }}

\noindent{\bf Parity:}
Like in the cyclic case of Ref.\DFTS,
the components of $\Psi$ have a definite parity as polynomials of $\tau$. As this parity
is reversed by each action of $\Delta_i$ (i.e. each action of $e_i$ on the link patterns)
we may define unambiguously a sign $\epsilon(\pi)$ for each link pattern $\pi$, with the
boundary condition that the ``maximal component" with link pattern $\pi_{max}$ connecting points
$2i-1$ to $2i$ (with the last point unmatched in odd size) has sign $\epsilon(\pi_{max})=1$.
Then $\Psi_\pi(-\tau)=\epsilon(\pi) \Psi_\pi(\tau)$ for all $\pi$. Note that with this definition
we also have
\eqn\defparity{
\epsilon(\pi)=(-1)^{\beta(\pi)}}
for all $L$, as a consequence of \valdeg.

\noindent{\bf Integrality, symmetry:}
All entries of $\Psi(\tau)$ are polynomials with non-negative integer coefficients,
and so are the sum rules $\Pi_L(\tau)$. We note that the entries of $\Psi$ are
not in general symmetric under reflection $\pi\to \pi^t$ of link patterns with 
respect to a vertical axis, namely
$\Psi_\pi(\tau)\neq \Psi_{\pi^t}(\tau)$ in general, although this symmetry is restored
at the RS point, where $\Psi_\pi(1)=\Psi_{\pi^t}(1)$. This is because the boundary conditions
on the left and right are {\it not} the same in general: $\Psi$ is indeed symmetric under 
$z_1\to 1/z_1$ on the left and $z_L\to q^6/z_L$ on the right (up to multiplicative factors $c_1$
or $c_L$, see Eq.~\sysopqkz), and the left-right symmetry
is restored only when $q^6=1$ (thus including the RS point $q=-e^{i\pi/3}$, $\tau=1$ and its 
``conjugate" $q=e^{i\pi/3}$, $\tau=-1$).

\newsec{Plane Partitions with symmetries}

In this section, we recall a few know facts on Plane Partitions with symmetries, 
related in particular to their expression as NILP and to their explicit enumeration.
We also introduce a generalization of CSTCPPs to reproduce the number of VSASMs.

\subsec{TSSCPPs and a first conjecture}

\fig{A sample NILP in bijection with a TSSCPP of size $10$. The corresponding endpoints
are circled, and read $r_1=1,r_2=3,r_3=4,r_4=7$.}{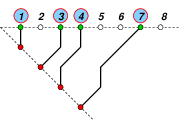}{6.cm}
\figlabel\tsscpp

We recall the expression for the number $N_{10}(2n)$ of TSSCPPs in a box 
of size $2n\times 2n\times 2n$. We refer for instance to \Bre\ for details and
further references. 
As TSSCPPs are maximally symmetric rhombus tilings of a regular hexagon of the triangular lattice
of size $2n$, they are entirely determined by the tiling of a fundamental domain of area
$1\over 12$th of the hexagon, with the shape of a triangle. Following the sequences 
of two of the three types of rhombi used, one easily ends up with an equivalent configuration
of $n-1$ NILP drawn on the integer plane (see Fig.\tsscpp\ for an illustration for $n=5$), 
starting at points $(i,-i)$, $i=1,2,\ldots,n-1$
and ending on the line $y=0$, making only vertical $(0,1)$ or diagonal steps $(1,1)$. We
record by an increasing sequence $1\leq r_1<r_2<\cdots <r_{n-1}$ the endpoints $(r_i,0)$.
Note that $r_i\leq 2i$, as the largest $r$'s are attained by using only diagonal steps.

The total number of TSSCPPs of size $2n$ equals
\eqn\tsscpp{ N_{10}(2n)=\sum_{1\leq r_1<r_2<\ldots <r_{n-1}} \det_{1\leq i,j\leq n-1}\left(
{i\choose r_j-i}\right)  =1,2,7,42,429,\ldots}
for $n=1,2,3,4,5\ldots$ 
The latter is expressed as the sum of minors of size $n-1$
of the $(n-1)\times (2n-2)$ matrix $Q$ with entries $Q_{i,r}={i\choose r-i}$,
$1\leq i\leq n-1$, $1\leq r\leq 2n-2$. This is in fact a particular case of the Lindstr\"om
Gessel Viennot (LGV) formula \LGV, expressing the number of lattice paths with fixed origins and
endpoints as a (fermionic Slater) determinant.

\fig{The bijection between link patterns of size $L=2n$ and sequences of integers
$1\leq r_1<r_2<\cdots <r_{n-1}$ with $r_i\leq 2i$ for all $i$ is illustrated on
an example for $n=5$. Starting from the link pattern (a), we first reflect it
with respect to a vertical axis (b), and then record the positions (c) of all origins of semi-circles by 
the corresponding point label minus one, omitting the first 
(at position $0$). These read $r_1=2,r_2=3,r_3=4,r_4=7$ here.}{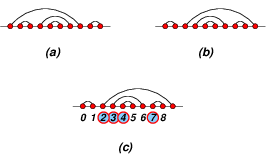}{12.cm}
\figlabel\endtolipa

In Ref.\DF, it was noted that the TSSCPPs may be regrouped (refined) according to their
common endpoints $\{r_1,r_2,\ldots,r_{n-1}\}$ into exactly $c_n$ sets, corresponding
to the conditions that $1\leq r_1<\cdots <r_{n-1}$ and $r_i\leq 2i$ for all $i$, where
the latter conditions ensure that the binomial coefficient ${i\choose r_i-i}={i\choose 2i-r_i}$
does not vanish. Listing the endpoints in lexicographic order, we may form
vectors $\Theta^{(2n)}$ with $c_n$ components, with entries equal to $\det_{1\leq i,j\leq n-1}
\Bigg({i\choose r_j-i}\Bigg)$, that sum to $N_{10}(2n)$. The first few such vectors read
\eqn\firfewtss{\eqalign{
\Theta^{(2)}&=\{1\}\cr
\Theta^{(4)}&=\{1,1\}\cr
\Theta^{(6)}&=\{1,2,1,2,1\}\cr
\Theta^{(8)}&=\{1,3,3,1,5,6,2,3,1,5,6,2,3,1\}\cr
}}
summing respectively to $1$, $2$, $7$, $42$. In Ref.\DF, a simple bijection between the set
of admissible endpoints and the link patterns was proposed. Here we use a slight modification
thereof, as we compose it with a reflection with respect to a vertical axis. 
This is summarized in Fig.\endtolipa.
Starting from a link pattern $\pi$ of size $L=2n$, we first reflect it with 
respect to a vertical axis, and
then record the positions of all origins of semi-circles forming it, by the point label minus one,
omitting the first one. This gives a bijective mapping $\pi\to \{r_i(\pi)\}_{1\leq i\leq n-1}$. Conversely,
given the $r_i$'s, there is a unique link pattern $\{r_1,\ldots,r_{n-1}\}\to \pi(r_1,\ldots,r_{n-1})$.
In Ref.\DFTS, the TSSCPPs were enumerated with a weight $\tau$ per vertical step, resulting
in generating polynomials
\eqn\tssctau{N_{10}(2n;\tau)=\sum_{1\leq r_1<r_2<\ldots <r_{n-1}} \det_{1\leq i,j\leq n-1}
\left(\tau^{2i-r_j}{i\choose r_j-i}\right)}
The latter were then conjectured to match the sum rules for the suitably normalized cyclic boundary
qKZ solutions. This $\tau$-enumeration leads naturally to the vectors $\Theta^{(2n)}(\tau)$, the entries of
which count the TSSCPPs with fixed endpoints (still listed in lexicographic order)
and with a weight $\tau$ per vertical step, summing
to $N_{10}(2n;\tau)$. The first few of them read
\eqn\fifrem{\eqalign{
\Theta^{(2)}(\tau)&=\{1\}\cr
\Theta^{(4)}(\tau)&=\{\tau,1\}\cr
\Theta^{(6)}(\tau)&=\{\tau^3,2\tau^2,\tau,2\tau,1\}\cr
\Theta^{(8)}(\tau)&=\{\tau^6,3\tau^5,3\tau^4,\tau^3,5\tau^4,6\tau^3,
2\tau^2,3\tau^2,\tau,5\tau^3,6\tau^2,2\tau,3\tau,1\}\cr
}}

Let us now look at the qKZ solutions of Appendix A, corresponding to even sizes $L=2,4,6,8$. 
We note that the terms of smallest degree in $\tau$,
namely the valuation terms, coefficients of $\tau^{\beta(\pi)}$ in $\Psi_\pi(\tau)$,
produce exactly the entries of the vectors \firfewtss. More precisely, we have
\eqn\fircon{ \Theta^{(2n)}_{r_1,\ldots,r_{n-1}}(\tau)
=\Psi_{\pi(r_1,\ldots,r_{n-1})}(\tau)\Big\vert_{min}}
where the subscript $min$ stands for the lowest order term in $\tau$, 
(term $\tau^{\beta(\pi)}$ in $\Psi_\pi(\tau)$),
and this holds for $n=1,2,3,4$.
Note that indeed 
\eqn\denbeta{ \beta(\pi)=\sum_{i=1}^{n-1} 2i-r_i(\pi) }
produces the same power of $\tau$ both in $\Psi_\pi(\tau)\vert_{min}$ and in 
$\Theta^{(2n)}_{r_1(\pi),\ldots,r_{n-1}(\pi)}(\tau)$.
The same phenomenon is observed for the qKZ solution of odd size $L=2n-1$, 
with the obvious extension of the bijection between sets of admissible endpoints 
and link patterns of odd size. 
We are led to conjecture that the relation \fircon\ holds for arbitrary $n$, namely that
\eqn\trueconj{ \Psi_{\pi}(\tau)\sim  \det_{1\leq i,j\leq n-1} \left({i\choose
r_j(\pi)-i}\right) \times  \tau^{\beta(\pi)} \times \left\{ \matrix{ 1 & {\rm if} \ L=2n \cr
\tau^{u(\pi)-n} & {\rm if} \ L=2n-1 \cr} \right. }
when $\tau\to 0$.

As a concluding remark, recall the above observation that, in general, the entries of
$\Psi(\tau)$ are not symmetric under reflection of link patterns $\pi\to \pi^t$. This is
clearly the case for the conjectured leading $\tau\to 0$ term in $\Psi_\pi(\tau)$ \trueconj,
say for even $L=2n$. Indeed, the set
$\{2n-1-r_j(\pi^t)\}_{j=1}^{n-1}$ is the complement of the set $\{r_i(\pi)\}_{i=1}^{n-1}$
within $\{1,2,\ldots,2n-2\}$, hence the
coefficients \trueconj\ for $\pi$ and $\pi^t$ are in general distinct 
(although $\beta(\pi)=\beta(\pi^t)$). This lack of
symmetry in refined TSSCPPs is a puzzle when we compare the entries \firfewtss\ to those of the
cyclic qKZ solution at the RS point (counting FPLs), that are indeed reflection-symmetric.
It suggests the existence of a non-symmetric change of basis relating the vector
of FPL numbers to that of refined TSSCPPs.

\fig{The bijection between CSTCPPs and pairs of TSSCPPs. The CSTCPP at hand
is a tiling of a regular hexagon
of size $2n=8$ here, symmetric with respect to all axes passing through the 
center of the hexagon and the middle
of each edge, resulting in fixed rhombi (represented in red). We have delimited a fundamental 
domain (thick broken black line), which is further mapped onto a NILP configuration. 
The latter is cut into two halves,
each of which is identified with the NILP formulation 
of a TSSCPP of same size.}{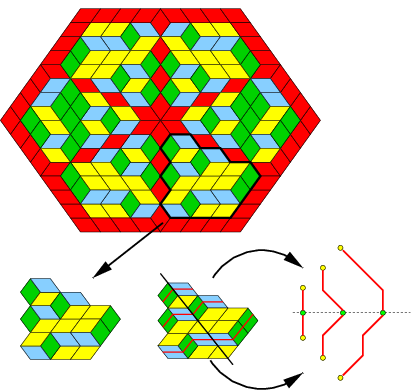}{13.cm}
\figlabel\bijcsts

\subsec{CSTCPPs and the case $L=2n-1$ }

The number of CSTCPPs in a regular hexagon of size 
$(2n)\times (2n)\times (2n)$, denoted by
$N_8(2n)$, was first obtained \MRR\ by mapping the corresponding rhombus tiling configurations to 
NILP, easily enumerated via a LGV-type determinant:
\eqn\cstcplgv{ N_8(2n)=\det_{1\leq i,j\leq n-1} \left({i+j\choose 2i-j}\right)}
The determinant was then evaluated as the product (see \CK\ for a simpler, illuminating proof):
\eqn\numcstcpp{ N_8(2n)= \prod_{i=0}^{n-1} (3i+1) {(6i)!(2i)!\over (4i+1)!(4i)!} =1,2,11,170,7429,...}
for $n=1,2,3,4,5,\ldots$

We may regard each CSTCPP as a {\it pair} of TSSCPPs. The bijection between CSTCPPs
and pairs of TSSCPPs is illustrated in Fig.\bijcsts. It is obtained by simply cutting
each CSTCPP into two halves, after rewriting it in terms of NILP.

This bijection results in the following identity, counting the total number of pairs 
of TSSCPPs with common arrival points $r_1<r_2<\cdots <r_{n-1}$:
\eqn\csscpp{N_8(2n)=\sum_{1\leq r_1<r_2<\ldots <r_{n-1} } 
\left(\det_{1\leq i,j\leq n-1}
\left({i\choose r_j-i}\right)\right)^2 }
One may derive this formula directly from \cstcplgv, by noting the following matrix identity:
denoting by $A$ and $B$ the matrices with entries $A_{i,r}={i\choose r-i}={i\choose 2i-r}$ and 
$B_{i,j}={i+j\choose 2i-j}$, with $i,j=1,2,\ldots n-1$ and $r=1,2,\ldots 2n-2$, we
have indeed that $B=AA^t$, as a consequence of the binomial identity
${i+j\choose 2i-j}=\sum_{r={\rm Max}(i,j)}^{{\rm Min}(2i,2j)} {i\choose 2i-r}{j\choose r-j}$.
Eq.\csscpp\ is nothing but a rewriting of the determinant of $B$ in terms of the minors of $A$.

\fig{A typical rhombus tiling of an hexagon of size $(2n)\times (2n+2)\times (2n)$ 
(with $n=4$ here) with a central
triangular hole of size $2\times 2\times 2$, symmetric with respect to all bissecting lines
of its edges. We have extracted a fundamental domain (thich black broken line) 
and transformed it into a configuration of NILP, by following successions of tiles of two 
of the three types used.
The latter are cut again into two different halves, one of which is a TSSCPP, the other
with paths of length one more.}{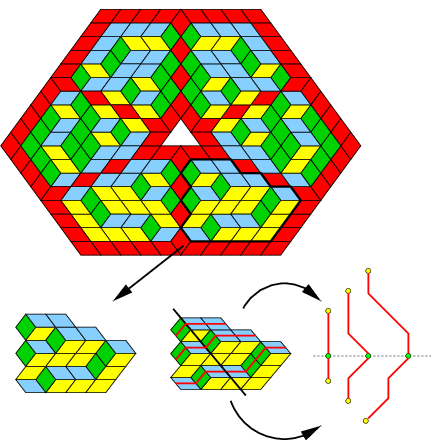}{12.cm}
\figlabel\newrhomb

\subsec{VSASMs and new Plane Partitions for the case $L=2n$ }

The total number $A_V(2n+1)$ of VSASMs of
size $(2n+1)\times (2n+1)$ equals:
\eqn\mnumvsasm{ A_V(2n+1)=\prod_{i=1}^n (3i-1){6i-3)!(2i-1)!\over (4i-1)!(4i-2)!} =1,3,26,646,45885,...}
for $n=1,2,3,4,5,\ldots$
It may also be expressed as a number of NILP, and actually reinterpreted as the number of rhombus tilings
of a hexagon of shape $(2n)\times (2n+2)\times (2n)$ with a
central triangular hole\foot{We refer to \CK\ and \CEKZ, for the weighted
enumeration of very similar objects,
also in relation with Descending Plane Partitions.} 
of size $2\times 2\times 2$ (see Fig. \newrhomb\ for an illustration).
The counting of NILP yields a determinant formula for the above numbers
\eqn\deterasm{ A_V(2n+1)=\det_{1\leq i,j \leq n-1} \left({i+j+1\choose 2i-j}\right) }
The latter determinant can be evaluated, using a more general result \MRR, leading to the
product formula \mnumvsasm.
As before, the NILP may be cut into two halves, one of which is identified with a TSSCPP,
and the other with a set of NILP of length one more (see Fig.\newrhomb). 

This bijection results in the following representation for the VSASM numbers:
\eqn\vsts{A_V(2n+1)=\sum_{1\leq r_1<r_2<\ldots <r_{n-1} }
\det_{1\leq i,j\leq n-1}\left({i\choose r_j-i}\right)\ 
\times \ \det_{1\leq i,j\leq n-1}\left({i+1\choose r_j-i}\right)}
which may also be derived from \deterasm\ via the binomial identity 
${i+j+1\choose 2i-j}=\sum_{r={\rm Max}(i-1,j)}^{{\rm Min}(2i,2j)} {i+1 \choose 2i-r}{j\choose r-j}$.

\newsec{$\tau$-enumeration of Plane Partitions with symmetries}

In this section we introduce polynomials $N_8(2n;\tau)$ and $A_V(2n+1;\tau)$ that generate
the rhombus tilings of Sects. 3.2 and 3.3 with a specific weighting by the parameter $\tau$.
These turn out to match the sum rule $\Pi_L(\tau)$ for the open qKZ solution, respectively
for $L=2n-1$ and $L=2n$ for all the examples of Appendix A, and we conjecture that they do in
general.

\subsec{Odd case $L=2n-1$}

We introduce the polynomials $N_8(2n;\tau)$ which enumerate the pairs of TSSCPPs involved in the 
CSTCPPs, with a weight $\tau$ per vertical step, except in the last step of say the second
TSSCP of the pair. The latter reads:
\eqn\taucss{\eqalign{N_8(2n;\tau)&=\sum_{1\leq r_1<r_2<\ldots <r_{n-1} } \det_{1\leq i,j\leq n-1}
\left(\tau^{2i-r_j}{i\choose r_j-i}\right)\cr
&\times \det_{1\leq i,j\leq n-1}
\left( \tau^{2i-r_j} {i-1\choose r_j-i-1}+ \tau^{2i-r_j-1} {i-1\choose r_j-i} \right) \cr}}
in which we have performed a decomposition of the paths of the second TSSCPP according to
their last step (receiving no weight $\tau$). 
Alternatively, the result \taucss\ may be put in the form of a single determinant, namely
\eqn\singdetcss{\eqalign{N_8(2n;\tau)&=
\det_{1\leq i,j\leq n-1}
\left( 
\sum_{r=1}^{2n-2}\tau^{2j-r}
{j\choose 2j-r} \right. \cr
&\qquad \times \left. \left\{ \tau^{2i-r} {i-1\choose 2i-r} +\tau^{2i-r-1}{i-1\choose 2i-r-1} \right\} 
\right)\cr}}

The first few polynomials $N_8(2n;\tau)$ take the values
\eqn\valtaucs{\eqalign{
N_8(2;\tau)&=1\cr
N_8(4;\tau)&=1+\tau\cr
N_8(6;\tau)&=1+3\tau+4\tau^2+2\tau^3+\tau^4\cr
N_8(8;\tau)&=1+6\tau+19\tau^2+32\tau^3+41\tau^4+35\tau^5+21\tau^6+11\tau^7+3\tau^8+\tau^9\cr
}}
Comparing with the data of Appendix A, we note that $\Pi_{2n-1}(\tau)=N_8(2n;\tau)$
for $n=1,2,3,4$. We also note the first few values of $N_8(2n;\tau)$ for $\tau=1,2,-1$:
\eqn\firfeneight{\eqalign{
N_8(2n;1)&=1,2,11,170,7429,920460,323801820,\ldots \cr
N_8(2n;2)&=1,3,55,6153,4196961,17446527483,441865841817751,\ldots \cr
N_8(2n;-1)&=1,0,1,0,81,0,456976,\ldots \cr}}
for $n=1,2,3,4,5,6,7,\ldots$ 
Apart from the obvious $N_8(2n;1)=N_8(2n)$, we have found the following identifications:
\eqn\identicss{\eqalign{
N_8(2n;2)&=\det_{1\leq i,j\leq n-1} \left({2i+2j-1\choose 2i-1}\right) \cr
N_8(2n;-1)&=\left\{\matrix{
A_V(2p+1)^4 & {\rm if}\  n=2p+1\cr
0 & {\rm if}\  n=2p\cr }
\right. \cr}}

The $\tau=2$ identification is proved as follows. In Ref.\DFTS, it was shown that the 
two rectangular $(n-1)\times (2n-2)$ matrices
$B$ and $A$ with respective entries $B_{i,r}={2i\choose r}$ and $A_{i,r}=2^{2i-r}{i\choose r-i}$,
$i=1,2,\ldots n-1$ and $r=1,2,\ldots 2n-2$,
actually share the same minors of size $n-1$, as one has $B=QA$, $Q$ the square matrix with 
entries $Q_{i,k}={k\choose i}$. 
Here we will use also a slight modification of this identity.
Introducing the matrices $C$, $D$ with entries $C_{i,r}={2i-1\choose r-1}$ and 
$D_{i,r}=2^{2i-r}{i-1\choose 2i-r}+2^{2i-r-1}{i-1\choose 2i-r-1}$, $i=1,2,\ldots n-1$ and
$r=1,2,\ldots 2n-2$, we have that $C=RD$ for a matrix $R$ with entries $R_{i,k}={i-1\choose k-1}$,
$i,k=1,2,\ldots n-1$. The determinant in \singdetcss\ at $\tau=2$ simply reads $\det(A D^t)$,
now reexpressed as $\det(QA D^tR^t)=\det(BC^t)$ as both square matrices 
$Q$ and $R^t$ have determinant $1$. But $BC^t$ has entries
\eqn\entribc{ (BC^t)_{i,j}=\sum_{r=1}^{2n-2} {2i\choose r}{2j-1\choose r-1}={2i+2j-1\choose 2i-1} }
and the first line of \identicss\ follows.

As to the $\tau=-1$ identification, we may recast \singdetcss\ at $\tau=-1$ as the determinant
$N_8(2n;-1)=\det(E)$ of a skew-symmetric matrix $E$, with entries 
$E_{i,j}= {i+j-1\choose 2i-j}-{i+j-1\choose 2i-j-1}$, thanks to standard binomial sum identites.
In Ref.\DFTS, as a consequence of a $\tau=-1$-enumeration of TSSCPPs,
a Pfaffian identity was conjectured for the square of the number of VSASMs, 
in the form
\eqn\sqpfa{ A_V(2n+1)^2=\pf_{1\leq i<j\leq 2n} \left(  \sum_{i\leq r<s\leq 2j}(-1)^{r+s-1}
\left\{ {i\choose r-i}{j\choose s-j}-{i\choose s-i}{j\choose r-j}\right\}\right) }
We simply note that
\eqn\notevs{ \eqalign{
\sum_{i\leq r<s\leq 2j}(-1)^{r+s-1}{i\choose r-i}{j\choose s-j}&={i+j-1\choose 2i-j}\cr
\sum_{i\leq r<s\leq 2j}(-1)^{r+s-1}{i\choose s-i}{j\choose r-j}&={i+j-1\choose 2i-j-1}={i+j-1\choose 2j-i}\cr}}
hence $\det(E)$ is nothing but the square of the Pfaffian \sqpfa, and the second line of 
\identicss\ boils down to the conjectured formula \sqpfa.

\subsec{Even case $L=2n$}

Like in the odd case, we introduce the polynomial $A_V(2n+1;\tau)$ that enumerates the 
rhombus tilings of the holed hexagon of Sect. 3.3, expressed as pairs of NILP, with a
weight $\tau$ per vertical step, except for the last steps of the second NILP of the pair.
It reads:

\eqn\tauvsasm{\eqalign{ A_V(2n+1;\tau)&=\sum_{1\leq r_1<r_2<\ldots <r_{n-1} } \det_{1\leq i,j\leq n-1}
\left(\tau^{2i-r_j}{i\choose r_j-i}\right)\cr 
&\times \det_{1\leq i,j\leq n-1}
\left( \tau^{2i-r_j+1} {i\choose r_j-i-1}+ \tau^{2i-r_j} {i\choose r_j-i}\right) \cr}}
Alternatively, this may be recast into a single determinant
\eqn\recvsa{\eqalign{
A_V(2n+1;\tau)&=\det_{1\leq i,j\leq n-1}\left(  \sum_{r=1}^{2n-2} \tau^{2i-r}{i\choose 2i-r}
\right.\cr
&\qquad \times  \left. \left\{ \tau^{2j-r+1}{j\choose 2j-r+1} 
+ \tau^{2j-r}{j\choose 2j-r}\right\}\right)\cr}}

The first few polynomials $A_V(2n+1;\tau)$ take the values:
\eqn\valvsasm{\eqalign{
A_V(3;\tau)&=1\cr
A_V(5;\tau)&=1+\tau+\tau^2\cr
A_V(7;\tau)&=1+3\tau+7\tau^2+6\tau^3+6\tau^4+2\tau^5+\tau^6\cr
A_V(9;\tau)&=1+6\tau+25\tau^2+54\tau^3+102\tau^4+119\tau^5+131\tau^6
+94\tau^{7}+67\tau^{8}+29\tau^{9}\cr
&+14\tau^{10}+3\tau^{11}+\tau^{12}\cr
}}
Comparing with the data of Appendix A, we note that $\Pi_{2n}(\tau)=A_V(2n+1;\tau)$
for $n=1,2,3,4$. We also note the first few values of $A_V(2n+1;\tau)$ for $\tau =1,2,-1$:
\eqn\fivalav{\eqalign{A_V(2n+1;1)&=1,3,26,646,45885,9304650,\ldots \cr
A_V(2n+1;2)&=1,7,307,82977,137460201,1392263902567,\ldots \cr
A_V(2n+1;-1)&=1,1,4,36,1089,81796,\ldots \cr}}
for $n=1,2,3,4,5,6,\ldots$
Apart from the obvious $A_V(2n+1;1)=A_V(2n+1)$, we have found the following identifications:
\eqn\idenvsa{\eqalign{A_V(2n+1;2)&=\det_{1\leq i,j \leq n} \left({2i+2j-3\choose 2i-1}\right) \cr
A_V(2n+1;-1)&=\left(N_8\Big(2\big[{n+1\over 2}\big]\Big)\, 
A_V\Big(2\big[{n\over 2}\big]+1\Big)\right)^2\cr}}
where $[x]$ stands for the integer part of $x$.

The first line of \idenvsa\ may be proved exactly by the same argument as before. As to the
second line, we note that at $\tau=-1$ \recvsa\ boils down to the determinant of a matrix $F$
with entries $F_{i,j}={i+j\choose 2i-j}-{i+j\choose 2i-j-1}$, $i,j=1,2,\ldots n-1$, 
thanks to standard binomial summation formulae. By simple row manipulations, we may
slightly transform $F$ as follows: let us introduce the matrix $P$
with entries $P_{i,j}=\delta_{i,j}+\delta_{i+1,j}$, $i,j=1,2,\ldots n-1$. Then
$FP$ has the entries 
$(FP)_{i,j}=\delta_{i,1}\delta_{j,1}+{i+j-1\choose 2i-j-2}-{i+j-1\choose 2j-i-2}$, 
$i,j=1,2,\ldots n-1$.
We note that when $n$ is odd, removing the first term $\delta_{i,1}\delta_{j,1}$ does not
change the value of the determinant, as the corresponding minor is that of a skew-symmetric
matrix of odd size ($n-2$), hence vanishes. Hence for $n=2p+1$, $\det(F)=\det(G)$, where $G$ is the
skew-symmetric matrix with entries $G_{i,j}= {i+j-1\choose 2j-i-2}-{i+j-1\choose 2i-j-2}$,
$i,j=1,2,\ldots 2p$.
Its determinant is therefore the square of its Pfaffian, which we conjecture to be
given by $\pf(G)=N_8(2p+2)A_V(2p+1)$. When $n$ is even, let us multiply the term $\delta_{i,1}\delta_{j,1}$
by some arbitrary real number $x$. Then the corresponding determinant takes the form $a x+b$,
as is readily seen by expanding it, say with respect to the first column. We have $b=0$ as it is
nothing but the determinant at $x=0$, in which case it is the determinant of a skew-symmetric
matrix of odd size. Finally $a$ is the $1,1$ minor. So at $x=1$, we get that 
for $n=2p$, $\det(F)=\det(H)$, where $H$ is the matrix with entries
$H_{i,j}= {i+j+1\choose 2j-i-1}-{i+j+1\choose 2i-j-1}$, $i,j=1,2,\ldots 2p-2$. The
determinant of this skew-symmetric matrix is the square of its Pfaffian, which we conjecture to be
equal to $\pf(H)=N_8(2p)A_V(2p+1)$.

\newsec{Conjectures and conclusion}

\subsec{Conjectures}

We list and comment the main conjectures of this paper, and add up a few.

\noindent{\bf Conjecture 1.} The coefficients of smallest degree in $\tau$ of the 
homogeneous open boundary qKZ solution
$\Psi^{(L)}(\tau)$ form a vector identical to that of TSSCPP numbers arranged according to 
their (admissible) endpoints, via the bijection with link patterns described in Sect. 3.1
(see Eq.~\trueconj).

This conjecture looks very promising, as it relates for the first time in a way similar to the
full RS conjecture two different objects, one of them purely combinatorial, here
the TSSCPPs arranged according
to their endpoints, and the other purely algebraic, in the form of the leading coefficients
of the qKZ solution $\Psi_\pi(\tau)$ when $\tau\to 0$. This remarkable coincidence suggests 
that the correspondingly refined TSSCPP numbers might be directly obtainable from the qKZ equation.
Moreover, if we were able to relate directly the cyclic and open boundary solutions of qKZ,
we would have a natural way of going from the components $\Psi(\tau)$ of the cyclic case,
equal presumably to some $\tau$-enumeration of ASMs or FPLs with fixed connectivities, 
to the TSSCPP numbers sorted according to their endpoints, which would provide us with a new
refinement in a possible TSSCPP-ASM correspondence.

\noindent{\bf Conjecture 2.} The sum rule for homogeneous open boundary qKZ 
solution $\Psi^{(L)}(\tau)$ equates the generating polynomial
for the corresponding Plane Partitions or rhombus tilings with the suitable reflection and
cyclic symmetries,
namely $N_8(L+1;\tau)$ if $L=2n-1$ and $A_V(L+1;\tau)$ if $L=2n$.

This produces a refinement of the sum rules proved in \DFOP, that incorporates the ``quantum"
deformation parameter $q$ explicitly.

\noindent{\bf Conjecture 3.} The "maximal" component $\Psi_{max}(L)$ of the qKZ solution in size $L$ 
with link pattern $\pi_{max}$ connecting points
$2i-1$ to $2i$, and the last point unmatched for odd size, reads respectively for even and odd sizes:
\eqn\maxcompeven{\eqalign{
\Psi_{\pi_{max}}(2n)&=\sum_{1\leq r_1<r_2<\ldots <r_{n-1} } \left(\det_{1\leq i,j\leq n-1}
\left(\tau^{2i-r_j}{i\choose r_j-i}\right)\right)^2 \cr}}
with values 
$1,1+\tau^2,1+5\tau^2+4\tau^4+\tau^6,1+14\tau^2+49\tau^4+62\tau^6+34\tau^8+9\tau^{10}+\tau^{12}$ 
for $n=1,2,3,4$.
\eqn\maxcompodd{\eqalign{
\Psi_{\pi_{max}}(2n-1)&=\sum_{1\leq r_1<r_2<\ldots <r_{n-1} } \det_{1\leq i,j\leq n-1}
\left(\tau^{2i-r_j}{i\choose 2i-r_j}\right)  \cr
&\qquad \times
\det_{1\leq i,j\leq n-1} \left(\tau^{2i-r_j-1}{i-1\choose 2i-r_j-1} \right) \cr}}
with values 
$1,\tau,2\tau^2+\tau^4,6\tau^3+13\tau^5+6\tau^7+\tau^9$
for $n=1,2,3,4$. 
The component of its reflected link pattern $\pi_{max}^t$, that 
leaves point $1$ un matched and connects all other points $2i$ to $2i+1$, $i\geq 1$, reads:
\eqn\maxcompoddsym{\eqalign{
\Psi_{\pi_{max}^t}(2n-1)&=\sum_{1\leq r_1<r_2<\ldots <r_{n-1} } \det_{1\leq i,j\leq n-1}
\left(\tau^{r_j-i}{i\choose r_j-i}\right) 
\det_{1\leq i,j\leq n-1} \left(\tau^{r_j-i}{i-1\choose r_j-i}\right)  \cr}}
with values 
$1,1,1+2\tau^2,1+6\tau^2+13\tau^4+6\tau^6$ for $n=1,2,3,4$. 

These are simply the complete $\tau$-enumeration of both CSTCPPs and VSASMs in the form of pairs of 
NILP, with a weight $\tau$ per vertical step in the two first cases and per diagonal 
step in the last one.

\noindent{\bf Conjecture 4.} In the open boundary case, the Razumov-Stroganov 
conjecture identifies the components 
$\Psi_\pi(2n)$ at $\tau=1$ with the numbers ${\rm VSFPL}_{2n+1}(\pi)$ of Vertically Symmetric 
Fully-Packed Loop configurations on a square grid
of size $2n+1$ reproducing the same connectivity pattern $\pi$. 
Combining this with our observation of Sect. 2.3 on the parity
of the components of $\Psi$ as polynomials of $\tau$ and with our $-1$-enumeration of VSASMs \idenvsa, 
we deduce a new alternating sum rule for the numbers ${\rm VSFPL}_{2n+1}(\pi)$: 
\eqn\altersum{ \sum_\pi \epsilon(\pi) \  {\rm VSFPL}_{2n+1}(\pi) 
=\left(N_8\Big(2\big[{n+1\over 2}\big]\Big)\, A_V\Big(2\big[{n\over 2}\big]+1\Big)\right)^2 }
with $\epsilon(\pi)$ given by \defparity.

This is the open boundary version of the alternating sum rule (5.1) of Ref.\DFTS.

\noindent{\bf Conjecture 5.} We have the two following identities for Pfaffians:
\eqn\conjfour{\eqalign{\pf_{1\leq i<j\leq 2p} \ \left({i+j-1\choose 2j-i-2}
-{i+j-1\choose 2i-j-2}\right)&=N_8(2p+2)\, A_V(2p+1)\cr
\pf_{1\leq i<j\leq 2p-2} \ \left({i+j+1\choose 2j-i-1}
-{i+j+1\choose 2i-j-1}\right)&=N_8(2p)\, A_V(2p+1)\cr}}

These have arisen from the $\tau=-1$-enumeration of VSASMs in the NILP form,
and should be compared with the determinantal expressions \cstcplgv\ and \deterasm\ 
respectively for the numbers $N_8(2n)$ and $A_V(2n+1)$. We suspect this is by far the easiest
to prove in the list of conjectures above, as both sequences $N_8$ and $A_V$ are known explicitly
and take simple product forms. This should presumably be done using techniques developed in \KK.

\subsec{Conclusion}

In this paper, we have found new conjectures giving a combinatorial interpretation of
the level one $U_q(sl_2)$ open qKZ minimal polynomial solution in terms of Plane
Partitions with various symmetries. The main interest is to have kept the dependence in
the quantum parameter $q$ via the combination $\tau=-(q+q^{-1})$, and to have related 
the qKZ solution to generating polynomials for the $\tau$-enumeration of Plane Partitions.
In particular, we have found a unifying framework for VSASMs and CSTCPPs, allowing to view both
as rhombus tilings of (possibly holed) hexagons with the same symmetries, and to 
$\tau$-enumerate them in a similar way.

The main and most promising conjecture regards the $\tau\to 0$ behavior of the components
of the solution, whose coefficients are interpreted as the numbers of TSSCPPs with fixed 
endpoints in their NILP formulation. This points to a possible approach of the 
ASM-TSSCPP correspondence, by trying to relate the cyclic and open qKZ solutions.
Note that no such nice coincidence seems to happen
in the cyclic case of \DFTS, but note however that the coefficients of top degree in $\tau$
coincide in the cyclic and open cases. This should not come as a surprise, as the 
$\tau\to \infty$ limit amounts to taking $q\to \infty$, and therefore leaves us
with the {\it same} renormalized operator $\Delta_i\to -z_i/(z_{i}-z_{i+1}) (\tau_i-1)$
and the {\it same} renormalized fundamental component 
$\Psi_{\pi_0}=z_1^{n-1}z_2^{n-2}\cdots z_{n-1}$
$z_{n+1}^{n-1}z_{n+2}^{n-2}\cdots z_{2n-1}$, for the open and cyclic cases. 

As it clearly appears from the studies of Ref.\DFTS\ and the present paper, there should
exist some sort of unifying interpretation of the minimal polynomial solutions of the
qKZ equation in terms of Plane Partitions or rhombus tilings. This should include also
the other boundary conditions considered in \ORBI\ as well as those with a point at infinity
along the cylinder \DFZJZ. The sum rules found in those cases so far do not have all plane partition
counterparts, but we believe such interpretations should always exist.

Another interesting question concerns the generalization to higher rank groups \ZJDF, where
sum rules again have produced nice integer sequences, without combinatorial interpretation
yet. Maybe one should hunt for some higher dimensional generalizations of Plane Partitions,
presumably with many symmetries.

Finally, let us comment on the specialization
$\tau=2$, corresponding to the rational limit $q\to -1$, known to produce
for the components of $\Psi$ the multidegrees of some variety of upper triangular nilpotent
matrices with additional symmetries \ORBI. Remarkably, we have obtained for the total
degree of these varieties (first lines of eqs.\identicss\ and \idenvsa)
the {\it same} total degree as that of the Brauer scheme of Refs.\DFZJb\ and \KZJ,
based on a completely different loop model with crossings, and moreover with periodic 
boundary conditions. This striking coincidence awaits a good geometrical explanation.
On the other hand, the polynomials $N_8(2n;2)$ and $A_V(2n+1;2)$ provide a nice
reexpression of the total degree of the Brauer scheme as a sum over CSTCPPs or their
even counterparts of powers of $2$, suggesting that, like in the cyclic case for
the variety $M^2=0$, these
Plane Partitions play the role of ``pipe dreams" \KM\ for the Brauer scheme, that would
be decomposable into complete intersections of linear and quadratic varieties.

\bigskip

\noindent{\bf Acknowledgments:}  This work was partially supported by the ENRAGE European network
MRTN-CT-2004-5616, the ANR program GIMP ANR-05-BLAN-0029-01, 
the ACI GEOCOMP and the ENIGMA European network MRT-CT-2004-5652.

\vfill\eject

\appendix{A}{Polynomial solution of the qKZ equation for open boundaries in the homogeneous limit}

\eqn\homopone{\eqalign{L=1\ \ \ \ \varphi_{\epsfxsize=0.4cm\vcenter{\hbox{\epsfbox{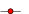}}}}&=1\cr
\Pi_1(\tau)&=1\cr}}
\eqn\homoptwo{\eqalign{L=2\ \ \ \ \varphi_{\epsfxsize=0.7cm\vcenter{\hbox{\epsfbox{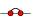}}}}&=1\cr
\Pi_2(\tau)&=1\cr}}
\eqn\homopthree{\eqalign{L=3\ \ \ \ 
&\varphi_{\epsfxsize=0.9cm\vcenter{\hbox{\epsfbox{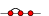}}}}=1\cr
&\varphi_{\epsfxsize=0.9cm\vcenter{\hbox{\epsfbox{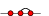}}}}=\tau\cr
\Pi_3(\tau)&=1+\tau\cr}}
\eqn\homopfour{\eqalign{L=4\ \ \ \ 
&\varphi_{\epsfxsize=1.2cm\vcenter{\hbox{\epsfbox{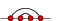}}}}=\tau\cr
&\varphi_{\epsfxsize=1.2cm\vcenter{\hbox{\epsfbox{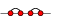}}}}=1+\tau^2\cr
\Pi_4(\tau)&=1+\tau+\tau^2\cr}}
\eqn\homopfive{\eqalign{L=5\ \ \ \ 
&\varphi_{\epsfxsize=1.5cm\vcenter{\hbox{\epsfbox{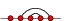}}}}=\tau\cr
&\varphi_{\epsfxsize=1.5cm\vcenter{\hbox{\epsfbox{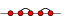}}}}=1+2\tau^2\cr
&\varphi_{\epsfxsize=1.5cm\vcenter{\hbox{\epsfbox{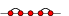}}}}=2\tau+\tau^3\cr
&\varphi_{\epsfxsize=1.5cm\vcenter{\hbox{\epsfbox{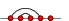}}}}=\tau^3\cr
&\varphi_{\epsfxsize=1.5cm\vcenter{\hbox{\epsfbox{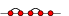}}}}=2 \tau^2+\tau^4\cr
\Pi_5(\tau)&=1+3\tau+4\tau^2+2\tau^3+\tau^4\cr}}
\eqn\homopsix{\eqalign{L=6\ \ \ \ 
&\varphi_{\epsfxsize=1.5cm\vcenter{\hbox{\epsfbox{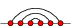}}}}=\tau^3\cr
&\varphi_{\epsfxsize=1.5cm\vcenter{\hbox{\epsfbox{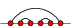}}}}=2\tau^2+2\tau^4\cr
&\varphi_{\epsfxsize=1.5cm\vcenter{\hbox{\epsfbox{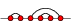}}}}=\tau+3\tau^3+\tau^5\cr
&\varphi_{\epsfxsize=1.5cm\vcenter{\hbox{\epsfbox{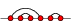}}}}=2\tau+2\tau^3+\tau^5\cr
&\varphi_{\epsfxsize=1.5cm\vcenter{\hbox{\epsfbox{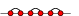}}}}=1+5\tau^2+4\tau^4+\tau^6\cr
\Pi_6(\tau)&=1+3\tau+7\tau^2+6\tau^3+6\tau^4+2\tau^5+\tau^6\cr}} 
\eqn\homopeight{\eqalign{L=7\ \ \ \ 
&\varphi_{\epsfxsize=1.8cm\vcenter{\hbox{\epsfbox{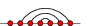}}}}
=\tau^3\cr
&\varphi_{\epsfxsize=1.8cm\vcenter{\hbox{\epsfbox{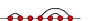}}}}
=3\tau^2+5\tau^4+\tau^6\cr
&\varphi_{\epsfxsize=1.8cm\vcenter{\hbox{\epsfbox{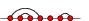}}}}
=5\tau^3+3\tau^5+\tau^7\cr
&\varphi_{\epsfxsize=1.8cm\vcenter{\hbox{\epsfbox{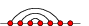}}}}
=\tau^6\cr
&\varphi_{\epsfxsize=1.8cm\vcenter{\hbox{\epsfbox{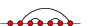}}}}
=2 \tau^2+3\tau^4\cr
&\varphi_{\epsfxsize=1.8cm\vcenter{\hbox{\epsfbox{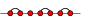}}}}
=3\tau+11\tau^3+10\tau^5+2\tau^7\cr
&\varphi_{\epsfxsize=1.8cm\vcenter{\hbox{\epsfbox{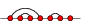}}}}
=5\tau^4+4\tau^6+\tau^8\cr
&\varphi_{\epsfxsize=1.8cm\vcenter{\hbox{\epsfbox{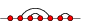}}}}
=2\tau+3\tau^3+3\tau^5\cr
&\varphi_{\epsfxsize=1.8cm\vcenter{\hbox{\epsfbox{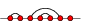}}}}
=3\tau^4+4\tau^6+\tau^8\cr
&\varphi_{\epsfxsize=1.8cm\vcenter{\hbox{\epsfbox{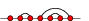}}}}
=\tau+6\tau^3+3\tau^5\cr
&\varphi_{\epsfxsize=1.8cm\vcenter{\hbox{\epsfbox{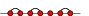}}}}
=6\tau^2+13\tau^4+6\tau^6+\tau^8\cr
&\varphi_{\epsfxsize=1.8cm\vcenter{\hbox{\epsfbox{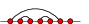}}}}
=3\tau^5+2\tau^7\cr
&\varphi_{\epsfxsize=1.8cm\vcenter{\hbox{\epsfbox{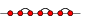}}}}
=1+8\tau^2+12\tau^4+5\tau^6\cr
&\varphi_{\epsfxsize=1.8cm\vcenter{\hbox{\epsfbox{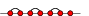}}}}
=6\tau^3+13\tau^5+6\tau^7+\tau^9\cr
\Pi_7(\tau)=1+6&\tau+19\tau^2+32\tau^3+41\tau^4+35\tau^5+21\tau^6
+11\tau^7+3\tau^8+\tau^9\cr}}
\eqn\homopeight{\eqalign{L=8\ \ \ \ 
&\varphi_{\epsfxsize=1.8cm\vcenter{\hbox{\epsfbox{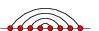}}}}
=\tau^6\cr
&\varphi_{\epsfxsize=1.8cm\vcenter{\hbox{\epsfbox{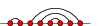}}}}
=\tau^3+6\tau^5+6\tau^7+\tau^9\cr
&\varphi_{\epsfxsize=1.8cm\vcenter{\hbox{\epsfbox{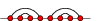}}}}
=3\tau^2+9\tau^4+12\tau^6+5\tau^8+\tau^{10}\cr
&\varphi_{\epsfxsize=1.8cm\vcenter{\hbox{\epsfbox{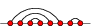}}}}
=5\tau^3+5\tau^5+3\tau^7+\tau^9\cr
&\varphi_{\epsfxsize=1.8cm\vcenter{\hbox{\epsfbox{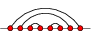}}}}
=3\tau^5+3\tau^7\cr
&\varphi_{\epsfxsize=1.8cm\vcenter{\hbox{\epsfbox{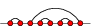}}}}
=2\tau^2+15\tau^4+24\tau^6+13\tau^8+2\tau^{10}\cr
&\varphi_{\epsfxsize=1.8cm\vcenter{\hbox{\epsfbox{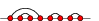}}}}
=3\tau+15\tau^3+29\tau^5+20\tau^7+7\tau^9+\tau^{11}\cr
&\varphi_{\epsfxsize=1.8cm\vcenter{\hbox{\epsfbox{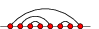}}}}
=5\tau^4+6\tau^6+3\tau^8\cr
&\varphi_{\epsfxsize=1.8cm\vcenter{\hbox{\epsfbox{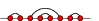}}}}
=2\tau+15\tau^3+27\tau^5+19\tau^7+7\tau^9+\tau^{11}\cr
&\varphi_{\epsfxsize=1.8cm\vcenter{\hbox{\epsfbox{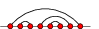}}}}
=3\tau^4+8\tau^6+3\tau^8\cr
&\varphi_{\epsfxsize=1.8cm\vcenter{\hbox{\epsfbox{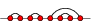}}}}
=\tau+12\tau^3+28\tau^5+25\tau^7+8\tau^9+\tau^{11}\cr
&\varphi_{\epsfxsize=1.8cm\vcenter{\hbox{\epsfbox{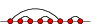}}}}
=6\tau^2+21\tau^4+18\tau^6+9\tau^8+2\tau^{10}\cr
&\varphi_{\epsfxsize=1.8cm\vcenter{\hbox{\epsfbox{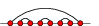}}}}
=6\tau^3+21\tau^5+18\tau^7+5\tau^9\cr
&\varphi_{\epsfxsize=1.8cm\vcenter{\hbox{\epsfbox{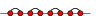}}}}
=1+14\tau^2+49\tau^4+62\tau^6+34\tau^8+9\tau^{10}+\tau^{12}\cr
\Pi_8(\tau)&=1+6\tau+25\tau^2+54\tau^3+102\tau^4+119\tau^5+131\tau^6
+94\tau^{7}\cr &+67\tau^{8}+29\tau^{9}+14\tau^{10}+3\tau^{11}+\tau^{12}\cr}}

\listrefs

\end